# Highly efficient BiVO$_4$ single-crystal nanosheets with dual modification: phosphorus doping and selective Ag modification

Can Fu[1,2,3], Baoyun Xu[3], Lingling Dong[3], Jinguo Zhai[3], Xuefei Wang[4*] and De-Yi Wang[1*]

[1]IMDEA Materials Institute, 28906 Getafe, Madrid, Spain
[2] E.T.S. de Ingenieros de Caminos, Universidad Politécnica de Madrid, 28040 Madrid, Spain
[3] State Key Laboratory of Polyolefins and Catalysis, Shanghai Engineering Research Center of Functional FR Materials, Shanghai Research Institute of Chemical Industry Co. LTD., Shanghai, 200062, China
[4]Department of Chemistry, School of Chemistry, Chemical Engineering and Life Sciences, Wuhan University of Technology, Wuhan 430070, People's Republic of China

E-mail: deyi.wang@imdea.org
E-mail: xuefei@whut.edu.cn

**Abstract**

BiVO$_4$, a visible-light response photocatalyst, has shown tremendous potential because of abundant raw material sources, good stability and low cost. There exist some limitations for further applicaitions due to poor capability to separate electron-hole pairs. In fact, a single-component modification strategy is barely adequate to obtain highy efficient photocatalytic performance. In this work, P substituted some of the V atoms from VO$_4$ oxoanions, namely P was doped into the V sites in the host lattice of BiVO$_4$ by a hydrothermal route. Meanwhile, Ag as an attractive and efficient electron-cocatalyst was selectively modified on the (010) facet of BiVO$_4$ nanosheets via facile photo-deposition. As a result, the obtained dually modified BiVO$_4$ sheets exhibited enhanced photocatalytic degradation property of methylene blue (MB). In detail, photocatalytic rate constant ($k$) was 2.285 min$^{-1}$g$^{-1}$, which was 2.78 times higher than pristine BiVO$_4$ nanosheets. Actually, P-doping favored the formation of O vacancies, led to more charge carriers, and facilitated photocatalytic reaction. On the other hand, metallic Ag loaded on (010) facet effectively transferred photogenerated electrons, which consequently helped electron-hole pairs separation. The present work may enlighten new thoughts for smart design and controllable synthesis of highly efficient photocatalytic materials.

## 1. Introduction

Energy-environmental catastrophes elicited via soaring population and hasty industrialization are the foremost tribulations of the past decade. In this regard, photocatalysts has evolved as a revolutionary and sustainable resolution [4-7, 52]. $BiVO_4$ as a visible-light response photocatalyst has shown tremendous potential for hydrogen evolution and degradation of organic pollutants because of abundant raw material sources, good stability, low cost, and narrow band gap [8, 9, 20, 36, 39]. However, pure $BiVO_4$ suffer some shortcomings such as: low charge-carrier transfermobility, short carrier diffusion length, and fast photoinduced carrier recombination.

Therefore, various modifications of $BiVO_4$ have been implemented. For example, doping a trace amount of impurity elements into the regular crystal lattice, which could change light absorptions or electrical properties of $BiVO_4$. In general, there are metallic (Ni, Fe, Mo) [34, 35, 37] and non-metallic (F, N, P) doping [23, 41]. Although having been proved to enhance the photocatalytic property of $BiVO_4$ effectively, in some cases metal doping exerted controversial influences. Metal dopant sometimes appeared as the center of recombination of electrons and holes in band structure [16]. Considering that, non-metal (F, N, P) doping is much more suitable for the modification of $BiVO_4$. N-doped and F-doped $BiVO_4$ have been extensively investigated [17, 23, 33, 38, 41, 42, 50]. On the contrary, few reports focused on the modifications of $BiVO_4$ with assistance of phosphorus doping [19, 47]. In fact, phosphorus doped photocatalysts have displayed superior photocatalytic activity in contrast to undoped ones [12, 26, 43].

Among various polymorphs of $BiVO_4$, monoclinic (ms-) $BiVO_4$ single-crystal nanosheets with exposed (010) facet exhibit the most efficient photocatalytic property [40, 49, 54, 56]. The fabrication of an internal electric field results in photogenerated carriers separation and transfer inside $BiVO_4$ [9, 20, 24, 39]. Recent reports proved that photocatalytic performance of $BiVO_4$ sheets highly relied on the exposure of (040) facet, which had higher charge carrier mobility and lower energy barrier than (110) facet [9, 22, 24, 27, 45]. It is believed that coupling crystal-face engineering with selective interfacial modifications for $BiVO_4$ nanosheets could be effective and efficient [27, 45]. Li et al. prepared oxidation co-catalyst ($PbO_2$, $MnO_x$, $Co_3O_4$) on the holes enriched (110) facet and reduction co-catalyst (Pt, Ag, Au) on the electrons enriched (040) facet, respectively. The results accelerated the charge separation, and dramatically further promoted the photocatalytic activity of $BiVO_4$ [24]. Our privious work also verified coupling crystal-facet engineering with selective interfacial noble-metal-modification was an effective approach to enhance photocatalytic property of $BiVO_4$ [25, 45]. For instance, $Ag_2O$-Ag was selectively modified on the (010) facet of single-crystal $BiVO_4$. The obtained $Ag_2O$-Ag/$BiVO_4$ exhibited enhanced photocatalytic decolorization rate of methyl orange with $6.8 \times 10^{-3}$ $min^{-1}$ [45]. Furthermore, Pt-Au/$BiVO_4$ photocatalyst with improved photocatalytic performace can be attributed to the synergistic effect of crystal-facet engineering and selective photodeposition of Pt-Au, which resulted in the orientation transport of photogenerated carriers in the single-crystal $BiVO_4$ [25].

Barely adequate is the obtained photocatalytic performance in a single-component modification strategy. Yu et al. found Eu and F co-doping gave rise to the synergetic effects: the increase in surface area and high separation efficiency of the photogenerated electrons and holes. Consequetly Eu/F-codoped $BiVO_4$ achieved higher photocatalytic activity for rhodamine B degradation reaction than Eu-doped $BiVO_4$ or F-doped $BiVO_4$ [50]. Moreover, Tan et al. constructed N-doped $BiVO_4$ single-crystalline nanoplates with exposed (040) facets with decomposition rate of Rhodamine B 97%. The result was about 2-fold enhanced compared with bare $BiVO_4$ (48%) [38]. Accounting for remarkable advantages of doping control and coupling crystal-facet engineering with selective interfacial deposition, the combination of these two strategies is expected to give full play to photocatalytic ability of $BiVO_4$ nanosheets.

In this work, $BiVO_4$ nanosheets were doped by phosphorus in a hydrothermal route with different contents of P, marked as BVP. Subsequently, metal Ag, as well-known efficient electron co-catalyst, was successfully deposited on the (010) facet of P-doped $BiVO_4$ sheets, marked as Ag/BVP. In this study, Ag only selectively modified the electron-rich (010) facet of P-doped $BiVO_4$ sheets by a photodeposition approach [24, 45]. The morphology, and microstructure of $BiVO_4$, BVP, Ag/BVP photocatalysts were systematically investigated by a series of characterization techniques. The photocatalytic performances of the synthesized samples were examined by the degradation of methylene blue (MB). To the best of our knowledge, here we firstly report that the enhanced photocatalytic activity of $BiVO_4$ sheets by a dual modification: (i) phosphorous doping which favored formation of O vacancies, increased concentration and mobility of charge carries and in turn promoted charge separation; (ii) coupling crystal-facet engineering with selective Ag modification on (010) facet which led to rapid orientation tansportation and catalytic reduction.

## 2. Experimental details

### 2.1 Materials

Bismuth (Ⅲ) nitrate pentahydrate ($Bi(NO_3)_3 \cdot 5H_2O$, $\geqslant$ 99.0%), ammonium metavanadate nitrate ($NH_4VO_3$, $\geqslant$

99.0%), silver nitrate (AgNO$_3$, ≥ 99.0%), methylene blue (C$_{16}$H$_{18}$ClN$_3$S), nitric acid (HNO$_3$, 65.0% - 68.0%), phosphoric acid (H$_3$PO$_4$ ≥ 85.0%), ammonia water (NH$_3$·H$_2$O, 25.0% - 28.0%), were all provided by Sigma-Aldrich Chemical Co.. All reagents were of analytical grade, and used without further purification. Deionized water was used in all experiments.

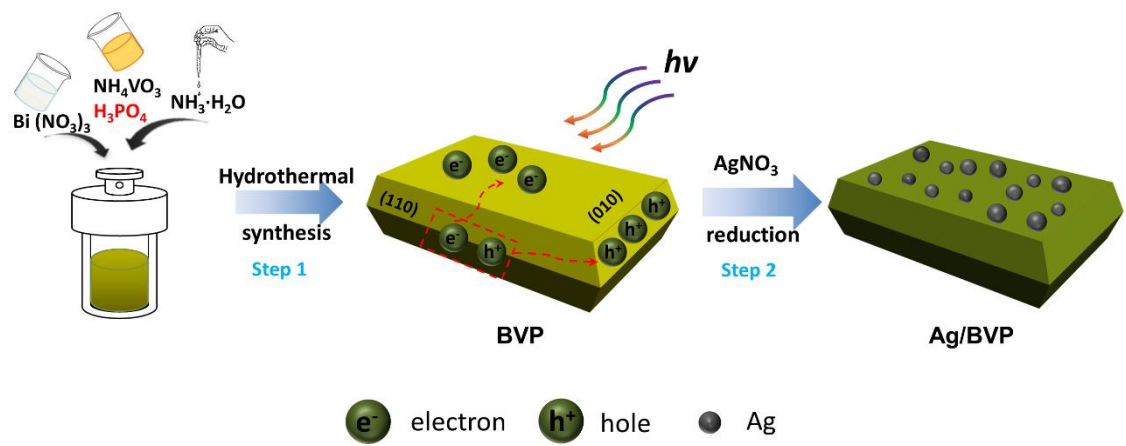

**Figure 1.** Scheme of synthetic route of dually modified BiVO$_4$ photocatalyst. Step (1): the hydrothermal preparation of P-doped BiVO$_4$ sheets with different ratio of P/(P+V); step (2): the photodeposition procedure for the selective modification of Ag nanoparticles on the (010) facets of P-doped BiVO$_4$ sheets.

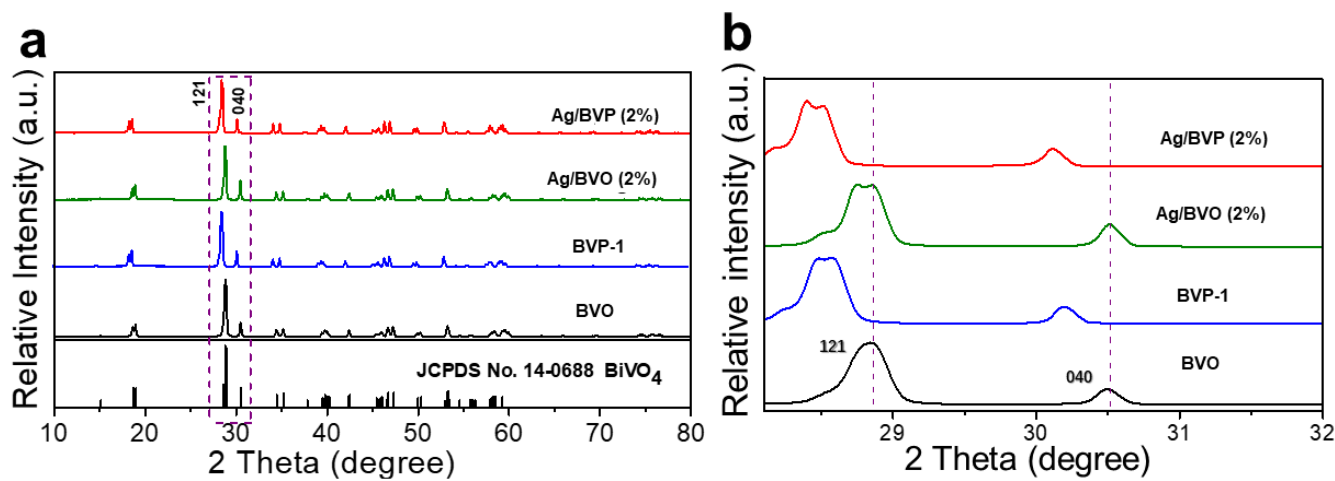

**Figure 2.** (a) Wide-angle XRD patterns of different samples, (b) magnified peaks of (121) and (040) planes.

## 2.2 Preparation of BiVO$_4$ (BVO) single-crystal nanosheets

As reported [24], Bi(NO$_3$)$_3$·5H$_2$O (7.2 mmol) and NH$_4$VO$_3$ (7.2 mmol) were dissolved in 60 mL HNO$_3$ solution (2 mol/L) and kept stirring for 1 h. After dissolving completely, the pH of solution were adjusted to 2.0 with adding NH$_3$·H$_2$O (25.0%-28.0%) by drops. After another 1 h stirring, the above collected yellow suspension was transferred to a Teflon-lined autoclave and put in a preheated oven at 180 ℃ for 24 h. The sample was centrifuged after the water and ethanol. After drying, yellow powders were collected finally, marked as BVO.

## 2.3 Preparation of P-doped BiVO$_4$ (BVP) sheets

The prepararion of BVP followed the same procedure as BiVO$_4$ nanosheets. V was substituted by P with different molar ratios (P/(P+V) = 0.5%, 1%, 2%, marked as BVP-0.5, BVP-1, BVP-2). In the preparation process, NH$_4$VO$_3$ was decreased with the calculated amounts as the corresponding addition of $H_3PO_4$. In detail, for (P/(P+V) = 0.5%), $Bi(NO_3)_3·5H_2O$ (7.2 mmol) and $NH_4VO_3$ (7.164 mmol) were dissolved in 60 mL $HNO_3$ solution (2 mol/L). Then 360 μL $H_3PO_4$ (0.1 mol/L) was added into above solution while stirring. After all chemicals dissolving completely, the pH of solution were adjusted to 2.0 with adding $NH_3·H_2O$ (25.0%-28.0%) by drops. The rest procedures were kept the same as the preparation of BVO with hydrothermal treatment. In the case of (P/(P+V) = 1% and 2%, the same procedures followed with different amount of $H_3PO_4$ and corresponding $NH_4VO_3$. Accoring to MB degradation test in this work (Figure 7(a)), BVP-1 exhibited highest photocatalytic activity among all the BVP samples. Therefore, in this case, the appropriate molar ratio of P/(P+V) was controlled to be 1%.

*2.4 Preparation of Ag/BVP sheets*

Ag/BVP sheets were prapared by a selective photodeposition approach [24]. Typically, 100 mg of the obtained BVP-1 powder from section 2.3 and calculated amounts of $AgNO_3$ solution (0.5 wt%, 1 wt%, 2 wt%) were stirred together in 100 mL of 10 vol% methanol solution, which marked as Ag/BVP (0.5%), Ag/BVP (1%), and Ag/BVP (2%). Keep the suspension stiring and bubbling with nitrogen in dark for 30 min. Subsequently, another 1 h stirring under Xe lamp irradiation (15 mW cm$^{-2}$), resultant powders in above suspension were collected by centrifugation. Based on the experimental data (Figure 7), Ag/BVP (2%) sample showed the best photocatalytic degradtion property. For comparison, 100 mg pristine $BiVO_4$ nanosheets were also selectively photodeposited Ag (2 wt%) on (010) facet, marked as Ag/BiVO (2%), which followed the same procedure with Ag/BVP (2%).

*2.5 Characterization*

Scanning electron microscope (SEM) analysis was executed on Zeiss Merlin Compact, Carl Zeiss. The X-ray photoelectron spectroscopy (XPS) analysis was conducted on a Thermo ESCALAB 250Xi spectrometer fitted with an X-ray source (Al Kα with photon energies 1486.6 eV). Raman spectra were tested on a DXR micro-spectroscopy system (Thermo Fisher Scientific Co., USA) by excitations with a 514.5 nm laser (45 mW for incident power). X-ray diffraction (XRD) patterns were measured by Philip XPERT-PRO diffractometer with Cu Kα radiation. The transmission electron microscopy (TEM, FEI Talos F200S, USA) with operated voltages of 200 kV and equipped with an X-Max 136 energy-dispersive X-ray spectrometer (EDS) were used to obtain TEM images and EDS data. In addition, the ultra-high resolution, high-angle annular dark-field scanning TEM (HAADF-STEM) and elements mapping analysis was further investigated microstructure information. UV–vis absorption spectra were obtained using a UV–vis spectrophotometer (UV-2450, Shimadzu, Japan). $BaSO_4$ was used as a reflectance standard in a UV–vis diffuse reflectance experiment. The Brunauer-Emmett-Teller (BET) specific surface area was calculated by the nitrogen adsorption-desorption method using ASAP 2460, micrometrics, automatic analyzer at 200 °C. The Inductively Coupled Plasma-Optical (ICP, Leiman, America) instrument was used to element analysis.

*2.6 Photocatalytic activity*

The evaluation of photocatalytic activity of the as-prepared samples was performed by the photocatalytic decolorization of methylene blue (MB) aqueous solution at ambien temperature. Typically, 0.05 g prepared sample was dispersed into 10 mL of MB solution (20 mg/L) while kept stirring in the whole process. Before irradiation, the suspension kept stiring for 2 h in the dark to ensure adsorption–desorption equilibrium. To evaluate UV-vis light photocatalytic activity, a 300 W xenon lamp (providing UV-vis light of 300 – 2000 nm) was used. The average light intensity which strikes the reaction solution surface was about 100 mW cm$^{-2}$. And the intensity changes of the absorption peak was measured at $\lambda$ = 664 nm to determine the concentration of the MB solution at certain time intervals by using a UV–vis spectrophotometer (759S, Shanghai Lengguang Technology Co. Ltd.).

For the evaluation of the effect of Ag loading, the MB degradation rate of Ag/BVP (2%) was futher tested under two different wavelenghths of LED lamps (600 and 480 nm, respectively). Two 600/480 nm LED lamps with an average light intensity of 80 mW cm$^{-2}$ (Shenzhen LAMPLIC Science Co. Ltd.) were used as light sources. The MB intensity measuement followed same steps as above.

Considering a low concentration of the MB aqueous solution, its photocatalytic decomposition of pollutes is a pseudo-first order reaction and its kinetics could be expressed as $\ln(c/c_0) = -kt$, where $k$ is the apparent rate constant, and $c_0$ and c are the concentrations of MB at initial state and after irradiation for t min, respectively [44].

## 3. Results and Discussion

*3.1 Strategy for the synthesis of Ag/BVP sheets*

The preparation strategy for the Ag/BVP sheets was followed by a facile two-step route in Figure 1. First, BVP sheets with (010)-dominant facets exposed were synthesized in a hydrothermal route by use of $Bi(NO_3)_3·5H_2O$, $NH_4VO_3$, $H_3PO_4$, and $NH_3·H_2O$ as precursors (Figure 1, step (1)). Second, it has been authenticated in many reports with spatial separation of photogenerated electrons and holes to the (010) and (110) facets of BiVO$_4$ sheets respectively [22, 24, 25, 45]. Due to different energy levels of two facets in the conduction bands and the valence bands, the electrons accumulation facet (0 1 0) and holes accumulation facet (1 1 0) prefer to reduction and oxidation reaction, respectively. Under this principle, Ag particles ought to be only photoreduced on the (0 1 0) crystal facet of BVP. In the whole photodeposition precess showed in step 2, AgNO$_3$ and methanol were employed as the source of Ag+ and hole scavenger respectively, which effectively separated the photo-generated electrons and holes. So the Ag/BVP sheets were controllably synthesized by the selectively photodeposition method.

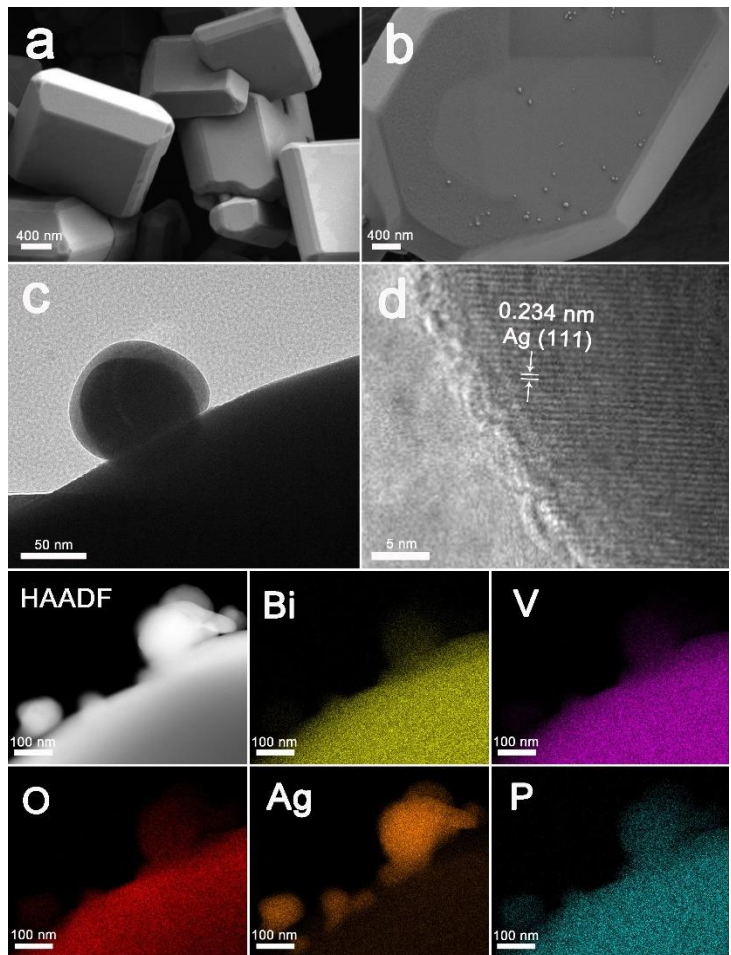

**Figure 3.** (a) SEM image of pristine BiVO$_4$ nanosheets, (b) SEM image and (c, d) HRTEM images of Ag/BVP (2%) and corresponding elemental mapping of Bi, V, O, Ag and P.

*3.2 Morphology and microstructure of Ag/BVP sheets*

According to the preparation part, we synthesized 4 types of photocatalysts: pristine BiVO$_4$ nanosheets, P-doped BiVO$_4$ sheets, selectively Ag loaded on (010) facet of P-doped BiVO$_4$ sheets, and selectively Ag loaded on (010) facet of pristine BiVO$_4$ sheets as reference. Based on the following photocatalytic degradation experimental datas, we picked BVO, BVP-1, Ag/BVP (2%), and Ag/BVO (2%) to do the comparation. In order to verify the phase structures of these samples, the XRD patterns are presented in Figure 2. From the wide-angle XRD curves in Figure 2(a), all the diffraction peaks of BVO, BVP-1, Ag/BVP (2%), and Ag/BVO (2%), matched well with the standard XRD pattern of BiVO$_4$ with a monoclinic crystal structure (JCPDS: No. 14-0688). Meanwhile, no other peaks were detected, it meant no other impurities existed during the synthesis. Notably, for BVP-1 and Ag/BVP (2%), the diffraction peaks slightly shifted to lower angle. The magnified patterns of (121) and (040) peaks in Figure 2(b) exhibited much more obvious shifts toward lower angles for BVP-1 and Ag/BVP (2%). This phonomenon may due to the change in the local crystal structures caused by P-doping, which as well as indicated that some amount of V$^{5+}$ has been well substituted by P$^{5+}$ without formation of any segregated impurities in the host BVO lattice [19, 32, 47, 53]. The diffraction peaks of Ag in BVO and Ag/BVP (2%) were not obvious by the XRD, which may be due to low content of Ag [11, 46, 54].

Table 1. ICP results of four samples.

| Sample | Element | wt % | Atomic % |
|---|---|---|---|
| BVO | Bi | 64.42 | 0.31 |
|  | V | 15.75 | 0.31 |
| BVP-1 | Bi | 76.80 | 0.37 |
|  | V | 16.36 | 0.32 |
|  | P | 0.02 | 0.0006 |
| Ag/BVO (2%) | Bi | 64.20 | 0.31 |
|  | V | 14.89 | 0.29 |
|  | Ag | 1.05 | 0.01 |
| Ag/BVP (2%) | Bi | 63.65 | 0.30 |
|  | V | 14.56 | 0.29 |
|  | Ag | 0.99 | 0.009 |
|  | P | 0.01 | 0.0003 |

The morphologies of BVO and Ag/BVP (2%) were observed by SEM in Figure 3 (a, b), respectively. BVO nanosheets were quite uniform with 400-600 nm thickness and 1-2 μm width. However, the image of Ag/BVP (2%) revealed that width of sheets enlarged to 2-7 μm at the same magnification although maintaining the similar sheets morphology of BVO, which was probably ascribed to preferred orientation growth along the (010) facet after P-doping [28]. Moreover, for Ag/BVP (2%), several Ag nanoparticles around 20–70 nm were verified to selectively deposited onto the (010) facet of BVP. Subsequently, the morphology and microstructure of the Ag/BVP (2%) sample were further examined by HRTEM (Figure 2c and d). Again, several nanoparticles around 20–70 nm could be clearly seen on the top of sheets in Figure 2 (c). Associating with the elemental mapping results for Ag/BVP (2%), phosphorous and silver elements were clearly proved to exist in this system. In particular, silver elements concentrated on the nanoparticles on the surface of sheets. Furthermore, inner lattice planes with spacing of ca. 0.234 nm in Figure 2 (d) were precisely determined to (111) planes of photodeposited metallic Ag. These observations is well coincided with the theoritical prediction in section 3.1. All of above demonstrated $BiVO_4$ single-crystal nanosheets with P-doping and selective Ag modification have been successfully synthesized.

The factual atomic ratios of P/V and percentage of Ag loading for four samples were examined by ICP analysis, respectively. Data in Table 1 exhibited that the actual concentrations of Ag and V of the BVP-1, Ag/BVO (2%) and Ag/BVP (2%) samples are less than initially introduced during the the process of synthesis. As reported, it seems the part of P/Ag not doped/deposited into/onto the lattice/surface of $BiVO_4$ has been dissolved and removed during the purification step.

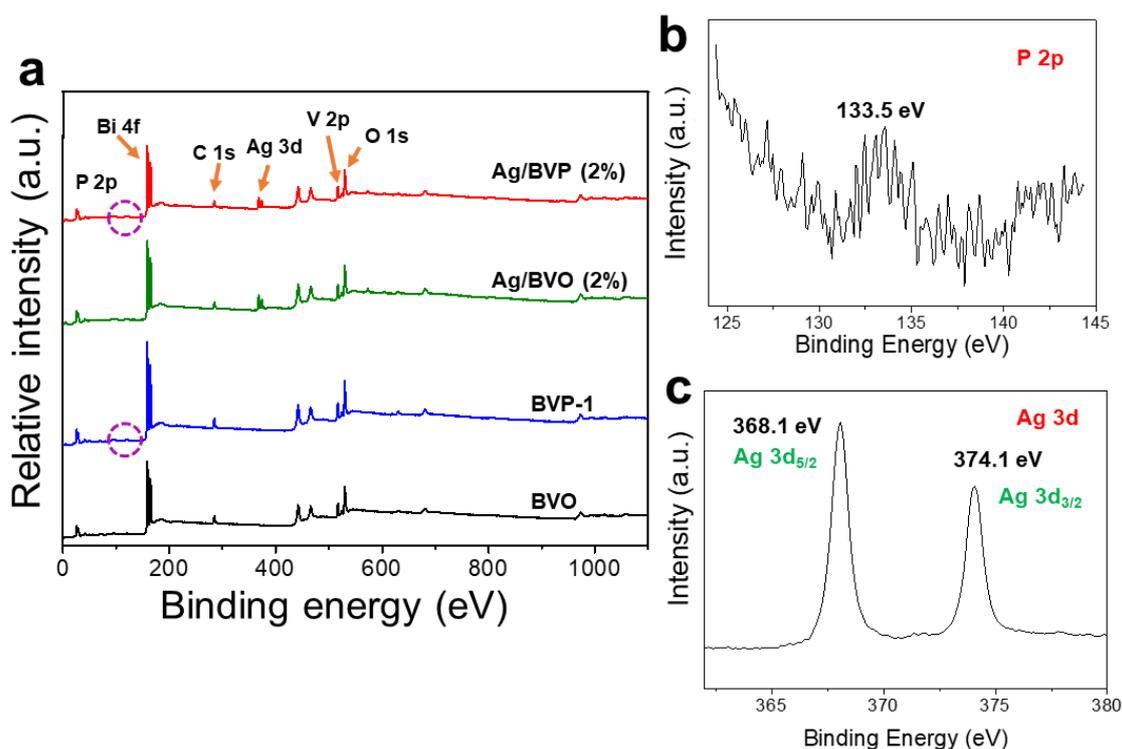

**Figure 4.** (a) Survey spectrum of XPS analysis for BVO, BVP-1, Ag/BVO (2%), Ag/BVP (2%), high-resolution XPS spectra for Ag/BVP (2%) sample (b) P 2p, and (c) Ag 3d.

In order to obtain accurate chemical composition and surface information, the samples were further analyzed by XPS. The survey XPS spectra was illustrated in Figure 4(a). The binding energy of C1s core level (284.6 eV) was refered to a standard. The co-existence of Bi, V, O, P, Ag elements were confirmed for Ag/BVP (2%) sample and no other impurities were detected in agreement with TEM and XRD results. Furthermore, the high-resolution spectra for Ag/BVP (2%) sample of P 2p, Ag 3d was displayed in Figure 4 (c, d) and Bi 4f, V 2p in Figure S1, respectively. The Bi 4f spectra in Figure S1 (a) showed that the peaks of Bi $4f_{5/2}$ and Bi $4f_{7/2}$ were assigned to 164.5 eV and 159.2 eV, respectively. The spin-orbit separation of 5.3 eV indicated the the obeservation of $Bi^{3+}$ ion [21]. The peak locations at 524.4, and 516.8 eV were associated to V $2p_{1/2}$ and V $2p_{3/2}$ orbitals, which was ascribed to $V^{5+}$ peaks in monoclinic $BiVO_4$ [13, 22]. The peak at 133.5 eV was assigned to P 2p corresponding to $P^{5+}$ in the phosphate in Figure 4(b) [48]. Besides, in Figure 4(c), the peaks at 374.1 and 368.1 eV were assigned to Ag $3d_{3/2}$ and Ag $3d_{5/2}$, which originated from $Ag^0$ [45]. Notably, each of the asymmetrical O 1s spectra could be deconvoluted into two peaks in Figure S2. The component with the lower value of binding energy was attributed to the lattice oxygen species, while the one with the higher binding energy value was ascribed to the chemisorbed oxygen species [18, 29, 32, 34]. In detail, peaks at 530.2 eV and 531.1 eV were assigned to O1s for BVO. Peaks at 530.0 eV and 530.9 eV were for Ag/BVP (2%). The lattice oxygen was assigned to the O atom in the $[Bi_2O_2]^{2+}$, while the adsorbed oxygen species were mainly $O^-$, $O^{2-}$ or $O_2^{2-}$ species dwelling at the O vacancies of the samples.[18, 49] Accorrding to the XPS results, the molar ratios of the chemisorbed oxygen to latice oxygen are 0.340 for BVO and 0.583 for Ag/BVP (2%), respectively. The amount of chemisorbed oxygen in Ag/BVP (2%) was much higher than BVO, indicating the concentration of O vacancies increased after dually modification.

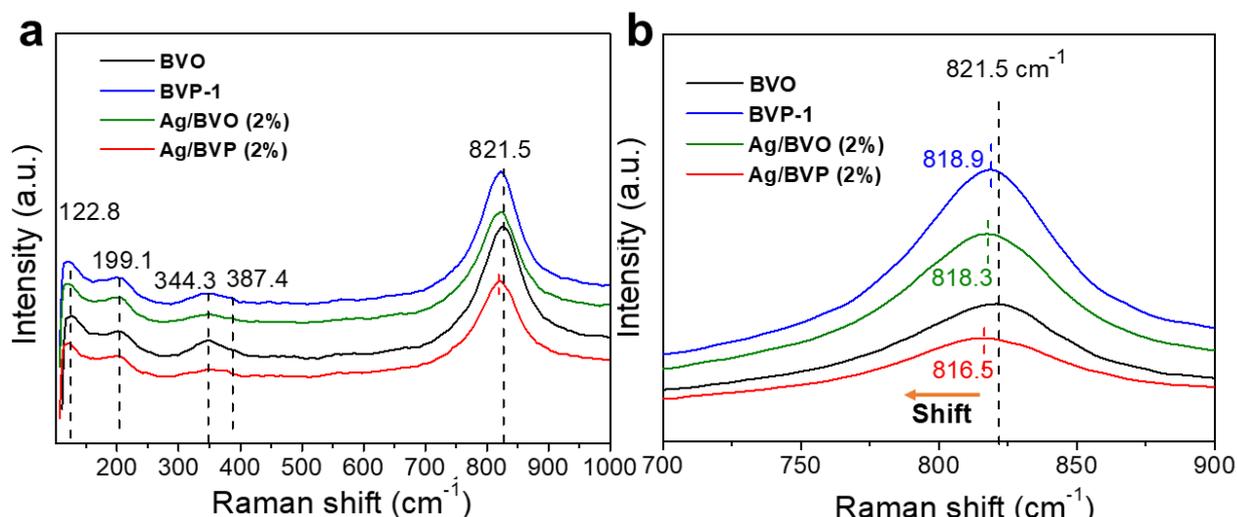

**Figure 5.** (a) Raman spectra of for BVO, BVP-1, Ag/BVO (2%), and Ag/BVP (2%) samples, (b) enlarge spectra of V-O stretching.

Raman spectra was displayed in Figure 5, which could characterize local structures and bonding states of samples. The peaks around 821.5, 387.4, 343.3, 199.1 and 122.8 cm$^{-1}$ were observed are the typical vibrational bands of ms-BiVO$_4$ in Figure 5(a) [14, 32]. The 122.8 cm$^{-1}$ and 199.1 cm$^{-1}$ bands are the rotation/translation modes of BiVO$_4$, whereas the bands at 343.3 cm$^{-1}$ and 387.4 cm$^{-1}$ are ascribed to the asymmetric and symmetric deformation modes of the VO$_4^{-3}$ tetrahedron, respectively [18]. As we know, the symmetric V-O stretching mode was located around 821.5 cm$^{-1}$, corresponding to the pronounced characteristic vibrational band for BiVO$_4$ [14, 27]. Obviously, this intense Raman band was induced to shift toward a lower frequency, no matter P-doping or selective modification of Ag particles to BiVO$_4$. The lower frequenc of the stretching band, the longer V-O bond length [51]. For BVP, the charge carriers increased after P-doping, which allowed for localization of electrons at the V center in VO$_4$. As a result, the V center was probably reduced, which led to the elongation of V-O bond in turn [30]. For Ag/BVO (2%) sample, the incorporation of Ag and BiVO$_4$ substrate may be responsible for the shift of the stretching band [43]. Therefore, after the dual modification of BiVO$_4$, Ag/BVP (2%) shifted most to 816.5 cm$^{-1}$. The above XPS and Raman experimental results strongly favored the dual modification of P-doping and coupling crystal-facet engineering with selective Ag depositon for BiVO$_4$ sheets were successfully achieved.

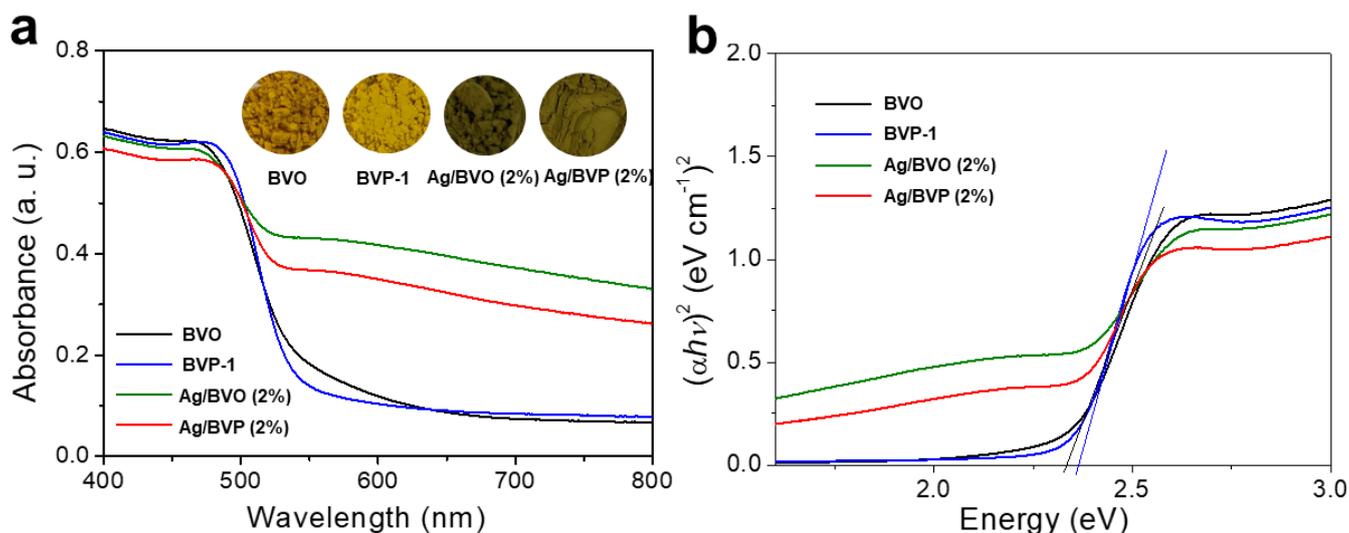

**Figure 6.** (a) UV-vis spectra and the corresponding digital images (inset), (b) the Kubelka-Munk plot for samples: BVO, BVP-1, Ag/BVO (2%), and Ag/BVP (2%).

The UV/vis absorption spectra of the corresponding digital images of BVO, BVP-1, Ag/BVO (2%), Ag/BVP (2%) samples were displayed in Figure 6. It was obvious that the BVP-1 exhibited a similar shape of UV-vis spectra in contrast of BVO owing to a very limited concrntration of P-doping. But Ag/BVO (2%) and Ag/BVP (2%) samples showed very different peak intensities and positions with additional shoulder peak between the 550 and 650 nm compared with BVO and BVP-1. This mainly resulted from the localized surface plasmon resonance (LSPR) of metallic Ag nanoparticles, which effectively broadens the visible light response range of the $BiVO_4$ sheets and increases the utilization of light energy [1, 11, 22, 44, 46, 55]. The band-gap energies of BVO, BVP-1, samples were estimated by means of expression: $\alpha h\upsilon = A (h\upsilon - E_g)^n$. While $\alpha$, $h\upsilon$, A, and $E_g$ are respectively indicated to the absorption coefficient, the energy of photon, the constant, and the optical bandgap energy. The value of n was taken to be 2 in this system (direct transition n = 2 and indirect transition n = 1/2) [19, 32]. The values of band-gap energy were achieved from extrapolating the linear content of $(\alpha h\upsilon)^2$ vs. $h\upsilon$ curves in the Figure 6(b). So the values of bandgap energy of BVO and BVP-1 were ca. 2.33 eV and 2.37 eV, respectively. For Ag-loading sample (Ag/BVO (2%) and Ag/BVP(2%)), it was not appropriate to estimate bandgap energy by this way. On the basis of DFT calculations in other report, P-doping could slightly enlarge the bandgap of $BiVO_4$ [19], which is in agreement with the obeservation in Figure 6(b). Furthermore, the insert digital images exhibited that the color of samples varied much in Figure 6(a). The pristine $BiVO_4$ nanosheets (BVO) was brilliant yellow. BVP-1 tend to be slightly brighter than BVO. Ag/BVO (2%) and Ag/BVP (2%) turned dark green and deep brown, respectively. The above results further demonstrated that controlled synthesis of various modified $BiVO_4$ sheets.

It is observed from Figure S3(a) that BVO and Ag/BVP (2%) samples exhibited a type II isotherm with a $H_3$ hysteresis loop . From the BET test, surface areas are 369.8 $m^2/g$ (BVO), and 561.8 $m^2/g$ (Ag/BVP (2%)), respectively. The larger surface area of the Ag/BVP (2%) can provide more surface active sites for the adsorption of the reactant $O_2$ molecules, resulting in more effective photocatalytic process.

### 3.3 Photocatalytic performance and mechanism

The photocatalytic performances involved with different ratios of P-doping samples (BVP-0.5, BVP-1, BVP-2) and different loading percentages of Ag on (010) facet of P-doped $BiVO_4$ (Ag/BVP (0.5%), Ag/BVP (1%), Ag/BVP (2%), Ag/BVP (3%), Ag/BVP (5%), and Ag/BVP (8%)) were evaluated by photocatalytic degradation of MB aqueous solution which was shown in Figure 7. During the test, the same percentage of Ag was loaded on the (010) facet of pristine $BiVO_4$ sheets as reference (Ag/BVO (2%)). The concentration of MB solution slightly decreased at the beginning of 30 min in dark but kept a constant value even longer. So after 2 h stirring in dark, various photocatalysts with MB solution were irradiated by the lamp. The corresponding results were observed in the Figure 7.

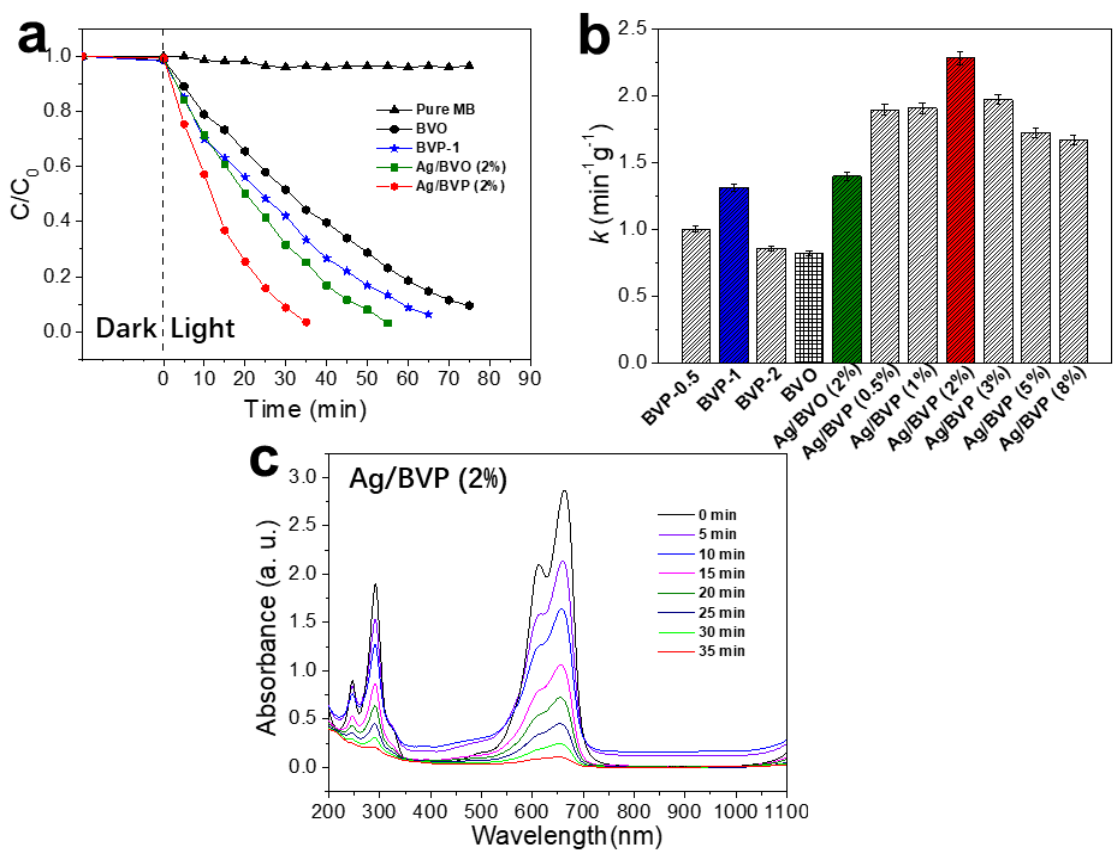

**Figure 7.** (a) Variations of C/C$_0$ of MB solution without and with synthesized photocatalysts under the irradiation, (b) rate constant (*k*) of MB decomposition by various synthesized photocatalysts, (c) absorbance spectra of MB solutions and its evolution after irradiation in the presence of Ag/BVP (2%).

When it comes to the series of P-doped BiVO$_4$, their rate constants (*k*) were 1.005 (BVP-0.5), 1.320 (BVP-1), and 0.859 min$^{-1}$g$^{-1}$ (BVP-2). The results demonstrated the photocatalytic activity was enhanced after P-doing in comparation of 0.823 min$^{-1}$g$^{-1}$ for pristine BiVO$_4$. A few calculations explored how P-doping improved photocatalytic activity of BiVO$_4$ [9, 12]. They reported P-doping could lead to the distortion of stucture and charge polarization, which was verified by Raman and XRD results. And more significantly, P-doping helped lower the energies of O vacancies formation, the concentration of O vacancies increased consequently. Generally speaking, an enhancement of catalytic activity of photocatalysts is associated with a higher oxygen vacancy density [18]. Othermore, Ag/BVP (0.5%), Ag/BVP (1%), Ag/BVP (2%) Ag/BVP (3%) Ag/BVP (5%) and Ag/BVP (8%) samples showed relatively high rate constants (*k*), which were 1.897, 1.911, 2.285, 1.975, 1.725, and 1.675 min$^{-1}$g$^{-1}$ in Figure 7. Samples of the selective Ag modificatiton on (010) facet of BVO and BVP showed much higher photocatalytic activity since coupling the crystal-facet engineering with efficient electron-cocatalyst Ag [45]. Specifically, rate constant of Ag/BVP (2%) increased about 2.78-fold compared to BVO, while 1.63, and 1.73 times higher than Ag/BVO (2%), BVP-1, respectively. Notably, When 2% of precious metal Ag is deposited on the BVP surface, the photocatalytic degradation efficiency of MB reached a maximum rate with *k* = 0.285 min$^{-1}$g$^{-1}$. With further increment of the loading percentage of Ag to 3, 5, 8%, photocatalytic efficiency decreased. Possibly excessive Ag deposition on the surface of the nanosheet reduced the reactive sites of BVO and increased the recombination probability of photogenerated carriers [3, 11]. The concentration of pure MB almost didn't show any reduction after 80 min irradiation, in Figure 7 (a). By analyzing the whole absorbance spectrum (200 – 1100 nm) of MB absorption in Figure 7(d), definitely no intermediates or other degradation products were detected during irradiation. Futhermore, the maximum absorption peak of MB at 664 nm did not drift [2, 15]. All above results demonstrated photodegradation of the MB molecule by these photocatalysts was actually occurring, and Ag/BVP (2%) exhibited highly efficient photocatalytic property.

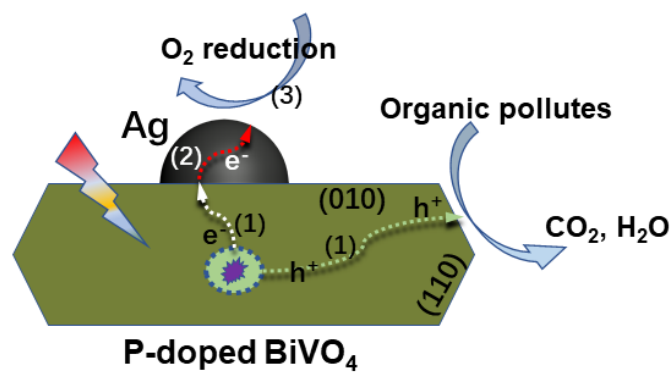

**Figure 8.** Possible photocatalytic mechanism of Ag/BVP (2%) photocatlyst: (1) orientation tansportation of photogenerated carriers; (2) interfacial transferring and (3) interfacial catalytic reaction.

Based on the above photocatalytic degradation results and analysis, it is very worthwhile to propose the potential photocatalytic mechanism of Ag/BVP (2%) photocatalyst, which was illustrated in Figure 8. P-doping possibly induced preferred orientation growth along the (010) facet of $BiVO_4$ sheets and consequent enlargement of (010) facet [28], facilitating further deposition of Ag. More significantly, the amount of O vacancy increased which was verified in O1s spectra (Figure S2), leding to more charge carriers, also offering higher mobility of carriers. In detail, some computational studies revealed that P doping could stabilize O vacancy structures by forming a defect complex, leading to higher concentrations of O vacancy. Subsequently, more O vacancies improved reducibility of the photocatalyst, which, in turn, increased the amount of charge carriers and increased the electronic conductivity, which could lead to superior photocatalytic activity [19, 30, 32, 47]. Other reports argued about the locations of O vancancy determining their functions [10, 29, 31]. O vacancy inside $BiVO_4$ acted as charge recombination centers, while O vacancy located on the surface could serve as good charge carrier traps as well as excellent adsorption sites. Charges were superior to concentrate around O vancany where rapidly transferring to adsorbed species can prevent electrons and holes recombination.

On the other hand, crystal-facet engineering of $BiVO_4$ meant the construction of $BiVO_4$ single crystals nanosheets with exposed (010)- dominant facet were achieved. After light irradiation, the photogenerated electrons and holes are spatially separated into corresponding (010) facet and (110) facet, seen in Figure 8(1). It was advantageous for this orientation transport of its photogenerated carries. Meanwhile selective modification of Ag on (010) facet of $BiVO_4$ boosted the interfacial transportation as Ag cocatalyst could extract the electrons rapidly from the conduction band of $BiVO_4$ and accelerate the catalytic reaction, seen in Figure 8 (2)(3).

In order to to illustrate the effect of Ag, the MB degradation rate of Ag/BVP (2%) was futher tested under two different wavelenghths of LED lamps (600 and 480 nm, respectively). In Figure 6(a), BVO showed an absorption edge near 550 nm, while Ag/BVP (2%) exhibited additional shoulder peak between the 550 and 650 nm result from LSPR of metallic Ag nanoparticles. 480 nm monochrome light mainly induced the band-gap excitation of BiVO$_4$, and 600 nm monochrome light only activated the LSPR of Ag nanoparticles. Under irradiation with 480 nm monochrome light, Ag/BVP (2%) maintained a high photocatalytic activity with *k* value of 0.230 min$^{-1}$g$^{-1}$, seen in Figure S4. Being irradiated by the 480 nm monochrome light, the photocatalytic activity of Ag/BVP (2%) decreased dramatically (*k* = 0.062 min$^{-1}$g$^{-1}$) . The above results strongly demonstrated that Ag particles here in Ag/BVP (2%) mainly functioned as an excellent cocatalyst while BVO played a major role in the photocatalysis, which was well confirmed by our previous work [45].

Based on the previous work we have done and literatures reported [1, 25, 45, 46, 55], metallic Ag cocatalyst in this system worked with two function: (1) Ag acted as excellent solid state electron mediator. In detail, the position of conduction band of BiVO$_4$ is more positive (0.03 V *vs.* NHE) than the potential of •O$^{2-}$/O$_2$ (-0.046 V). The photogenerated electrons of BiVO$_4$ could not effectively reduce O$_2$. After Ag nanoparticles were selectively modified on the (0 1 0) facet of BiVO$_4$ nanosheets, photogenerated electron generated on conduction band (CB) of BiVO$_4$ would continuously transfer across the BiVO$_4$–Ag interface to Ag due to the higher work function of Ag, resulting in stronger oxygen reduction capacity of the photogenerated electrons in this system. The above result is consistent with the Schottky barrier theory. Consequently the recombination of the electrons and holes was effectively suppressed with enhanced photocatalytic performance. (2) LSPR effect of Ag could help absorb photons from incident light and increase the utilization of energy.

Therefore, dual modifications of phosphorous doping and coupling crystal-face engineering with selective Ag deposition on (010) facet of BiVO$_4$, increased concentration of O vacancy and numbers of charge carries (P-doping), simutaneously promoted rapid orientation tansportation and interfacial charge transfer (selective Ag deposition), consequently improved charge separation and catalytic reaction.

## 4. Conclusion

In summary, a novel Ag/BVP (2%) photocatlyst with dual modification of P-doping and coupling crystal-face engineering with selective Ag depositon was successfully synthesized. This material was easily prepared by a facile two-step route involved with the initial construction of P-doped BiVO$_4$ sheets and following surface loaded Ag cocatalyst. The results showed that Ag/BVP (2%) obrained dramatically higher photocatalytic activity than BVO, BVP-1, or Ag/BVO (2%). Based on experimental results, a synergistic mechanism of P-doping and coupling crystal-face engineering with selective deposition of Ag was proposed in light of its highly enhanced photocatalytic activity. Namely, with the assitance of P-doping, concentration and mobility of charge carries was soared. Simutaneously coupling crystal-face engineering with selective Ag modifications on (010) facet led to rapid orientation tansportation, and interfacial charge transfer, consequently resulted in highly efficient catalytic reaction. The present work may enlighten new thoughts for smart design and controllable synthesis of highly efficient photocatalytic materials.

**Acknowledgements**

This research was supported by the Natural Science Foundation of Science and Technology Commission of Shanghai Municipality (19ZR1423700). Ms. Can Fu would thank the financial support from China Scholarship Council (201506950020).